\documentclass[11pt]{article}
\usepackage[utf8]{inputenc}
\usepackage[english]{babel}
\usepackage{amsmath}
\usepackage{parskip}
\usepackage{dsfont}
\usepackage{enumitem}
\usepackage{graphicx}
\usepackage{caption}
\usepackage{float}
\usepackage{amssymb}
\usepackage{cuted}
\usepackage{lipsum}
\usepackage{abstract}
\numberwithin{equation}{section}
\usepackage[left=1.5 cm, right=1.5 cm , top= 1.5 cm, bottom= 3 cm]{geometry}
\usepackage[backend=biber, style=phys]{biblatex}
\usepackage{hyperref}
\bibliography{Bibliography}

\usepackage{authblk}
\hypersetup{
    colorlinks=true,
    linkcolor=blue,
    filecolor=blue,  
    citecolor= blue,
    urlcolor=blue
    }
\date{}

\bibliography{Bibliography.bib}

\title{First Principle Description of Plasma Expansion Using the Expanding Box Model.}

\author[1,*]{Sebastián Echeverría-Veas}
\author[1,*]{Pablo S. Moya}
\author[2,3]{Marian Lazar}
\author[2,4]{Stefaan Poedts}

\affil[1,*]{ Departamento de Física, Facultad de Ciencias, Universidad de Chile, Las Palmeras 3425, Ñuñoa, Santiago 7800003, Chile; $^*$s.echeverria@ug.uchile.cl (S. E-V), pablo.moya@uchile.cl (P.S.M).}
\affil[2]{Centre for mathematical Plasma Astrophysics, Dept.\ of Mathematics, KU Leuven, Celestijnenlaan 200B, B-3001 Leuven, Belgium.}
\affil[3]{Institute for Theoretical Physics IV, Faculty for Physics and Astronomy, Ruhr University Bochum, D-44780 Bochum, Germany.}
\affil[4]{ Institute of Physics, University of Maria Curie-Skłodowska, ul.\ Radziszewskiego 10, 20-031 Lublin, Poland.}

\begin{document}

\maketitle

\begin{abstract}
\vspace{1cm}
  Multi-scale modeling of expanding plasmas is crucial for understanding the dynamics and evolution of various astrophysical plasma systems such as the solar and stellar winds. In this context, the Expanding Box Model (EBM) provides a valuable framework to mimic plasma expansion in a non-inertial reference frame, co-moving with the expansion but in a box with a fixed volume, which is especially useful for numerical simulations. Here, fundamentally based on the Vlasov equation for magnetized plasmas and the EBM formalism for coordinates transformations, for the first time we develop a first-principles description of radially expanding plasmas in the EB frame. From this approach, we aim to fill the gap between simulations and theory at microscopic scales to model plasma expansion at the kinetic level. Our results show that expansion introduces non-trivial changes in the Vlasov equation (in the EB frame), especially affecting its conservative form through non-inertial forces purely related to the expansion. In order to test the consistency of the equations, we also provide integral moments of the modified Vlasov equation, obtaining the related expanding moments (i.e., continuity, momentum, and energy equations). Comparing our results with the literature, we obtain the same fluids equations (ideal-MHD), but starting from a first principles approach. We also obtained the tensorial form of the energy/pressure equation in the EB frame. These results show the consistency between the kinetic and MHD descriptions. Thus, the expanding Vlasov kinetic theory provides a novel framework to explore plasma physics at both micro and macroscopic scales in complex astrophysical scenarios.
\end{abstract}
\vspace{1cm}

\section{Introduction}

Astrophysical plasmas have been widely studied through different theories, simulations, and observations based on plasma physics knowledge. 
In this case, we have the chance but also face the major challenge of studying plasma dynamics in various contexts, from kinetic processes conditioned by the energy (velocity) distributions of plasma particles to the macro-physics of a hydrodynamic plasma; or from high energy fully relativistic manifestations of plasmas in quasars and AGN jets, to non-relativistic solar outflows filling the heliosphere and planetary environments \cite{Benz_2012, Vereshchagin_2017,Reames_1999, Draine_2010}. However, precisely these complex manifestations of the plasma lead to multiple problems, which must be addressed and solved according to the phenomena of interest in our analysis. We can thus mention the modeling of turbulence and instabilities, from those at the micro- (kinetic) scales associated with the anisotropy of charged particles to those characteristic of large-scale plasma systems in different astrophysical scenarios \cite{Melrose1986, Ng1996, Howes2008, Hillier2016}. It is also not trivial to decode the energy transfer mechanisms, i.e., between particles and electromagnetic fields, plasma heating, and particle acceleration \cite{Toptyghin1980, jones1991, hededal2004, howes2010}. Not only the intrinsic properties arouse interest, but also the fact that natural plasmas are not isolated systems, but in interaction with the environment that affects their evolution \cite{verheest1996, bliokh_1995, gliddon_1966}. 

Plasma clearly dominates most of the visible matter in the universe, but its properties vary in various astrophysical or space contexts. For instance, plasmas surrounding black holes (accretion disks), AGN jets \cite{Balbus_1998, abramowicz_2013, yang_2016, blandford_2019}, highly contrast with heliospheric plasmas, solar wind, and the close-to-earth environment \cite{alfven_1947, parker1958, smith_1975, bruno_2013}. 
Under any of these circumstances, plasma is not a static system, it expands, shrinks, and constantly changes its structure. All of these are expanding plasma systems, whose analysis may invoke both kinetic and magnetohydrodynamic (MHD) theories, but sometimes also relativistic plasma approaches \cite{lemaire_1971, marsch_2006, chew_1956, matthaeus_1982, modena_1995, boris_1970}. Theories and numerical modeling aiming to a better understanding of the plasma expansion and quantifying its effects, e.g., in the heliosphere but also other astrospheres in our Galaxy, are very complex and computationally highly demanding. For instance, the most immediate problem when computing plasma simulations is purely related to computer limitations, especially, memory limitation to study the expansion of a plasma parcel. 
If we couple the kinetic physics and the expansion, not only computational time but also the memory needed to explore the possible effects would be ineffective. Therefore, more effective and methodological frameworks should be applied when studying plasma expansion.

\textcite{Velli.1992} proposed and developed the Expanding Box Model (EBM), which allows to study of plasma expansion in a new system of reference. The main idea of this model came as an answer to the limited memory in computer simulations when studying the solar wind plasma expansion. It enables to study of the radial-spherical expansion through a Cartesian approximation with a non-trivial change of coordinates. In this context, consider a plasma parcel expanding in a static and inertial system of reference $S$. The EBM defines a new system of reference $S'$ co-moving with the plasma at constant velocity. In this new framework, through the change of coordinates, the plasma parcel becomes steady (non-expanding) and is moving along with it. For instance, from the $S$ system an observer will see the plasma expanding and going away, while from the new $S'$ system, the observer will move along with the parcel but will not notice the expansion (i.e., the volume of the plasma parcel/box is constant). The question that arises is how can the expansion be described in a non-expanding frame. Within this new description, the expansion is no longer a spatial property of the system. When transforming to the new frame, the expansion traduces in temporal variations rather than spatial. This idea has the computational advantage of studying the expansion in a co-moving and non-expanding framework. Also, this way we can expect to reduce computational memory limitations in specific studies of plasma expansion. On the other hand, the EBM has analytical advantages, as we can rewrite equations using this formalism with an explicit dependence on the expanding parameters.

Originally proposed as a Cartesian description of coordinates, EBM has improved and generalized to a more diverse description through the years and more recent research. On one hand, \textcite{Grappin.1996} applied this model using polar coordinates to study the solar wind expansion in the ecliptic plane. On the other hand, more recent research has generalized this model by considering an accelerated co-moving frame \cite{Tenerani_2017}, enabling to study of expansion in the accelerating regions of the solar wind close to the Sun. These upgrades demonstrate how flexible and diverse the study of the expansion using the EBM can be. Thus, we can adapt the equations and the model to the applications of interest, which are in general focused on studying plasma physics in the macroscopic regime through MHD description \cite{Velli.1992,grappin.1993, Grappin.1996}. It is possible to rebuild the MHD approach in the co-moving frame, from the well-known MHD equations and the EBM formalism, and apply, for instance, in hybrid simulations with kinetic ions and fluid electrons. Following these ideas, \cite{Liewer.2001} studied the role of the expansion when both kinetic and fluid populations are in the plasma \cite{Ofman.2011, Moya.2012}. More recent studies have focused on the expansion effects in the simulations at microscopic or kinetic scales \cite{Innocenti_2019, Innocenti.2020, Micera.2020, Micera.2021}. The results appear to motivate the utility and advantage when incorporating the expansion through the EBM, affecting density, velocity, or even the magnetic field profiles. Physical quantities are thus conditioned by the EBM, showing, for instance, the temperature or magnetic field decreases, in agreement with the observations. Moreover, expansion may also play an indirect role in the heath flux regulation through the excited instabilities \cite{Innocenti.2020}. Under this context, \textcite{Seough_2023} introduced the effect of expansion at the kinetic level to add wave-particle relaxation to the well-known double adiabatic equations in the EB formalism. They considered the well-known moment-based quasilinear kinetic theory and added the EBM temperature evolution to the equations.

As mentioned, even though there have been diverse successful applications of the EBM in theoretical and, mainly, numerical simulations, there is still no Vlasov theory describing plasma expansion from a first principles approach, allowing us to describe both micro and macroscopic plasma physics from the expanding kinetic equation developed in this manuscript. In this paper, for the first time, we present a novel first principles description for spatially expanding plasmas, which relies on the Expanding Box formalism. Through this description, we introduce and develop a new theoretical framework, fundamentally based on the (collisionless) Vlasov equation written in the co-moving/EB frame, aiming to fill the gap between theory and simulations, which is especially relevant for the description of plasmas at kinetic scales. In particular, we present the general considerations and mathematical formalism when a microscopic description for expanding plasma is needed. Based on the transformed Vlasov equation, general expressions for the principal moments of the velocity distribution can be derived, as well as general MHD equations in the EB framework. From these expressions, we can explicitly test the consistency between micro and macroscopic physics.
Despite the fact that EBM was initially proposed for the study of the expanding solar wind, the main idea of this model is entirely general. Indeed, as starting from the Vlasov-Maxwell system most plasma descriptions can be obtained, this new formalism may open new ways to study the plasma physics of any expanding system, from a local-to-Earth environment, such as the solar wind, or to wider astrophysical contexts, such as expanding relativistic plasma. These ideas are the stepping stones for further research in expanding systems at kinetic scales.

This paper is organized as follows. Section \ref{section 2} introduces the general formalism and definitions used when working with the EBM. In Section \ref{section 3} we derive the expanding (collisionless) Vlasov equation, which transforms the standard expression to the co-moving frame. The expanding moments, i.e., continuity, momentum, and energy equations, are obtained in Section \ref{section 4}, and in Section \ref{section 5} we develop ideal MHD equations to compare with the results already existing in the literature. The last section summarizes our results and identifies a series of implications for future applications.

\section{Expanding Box Model (EBM)}
\label{section 2}

\begin{figure}[t]
    \centering
    \includegraphics[width=8cm]{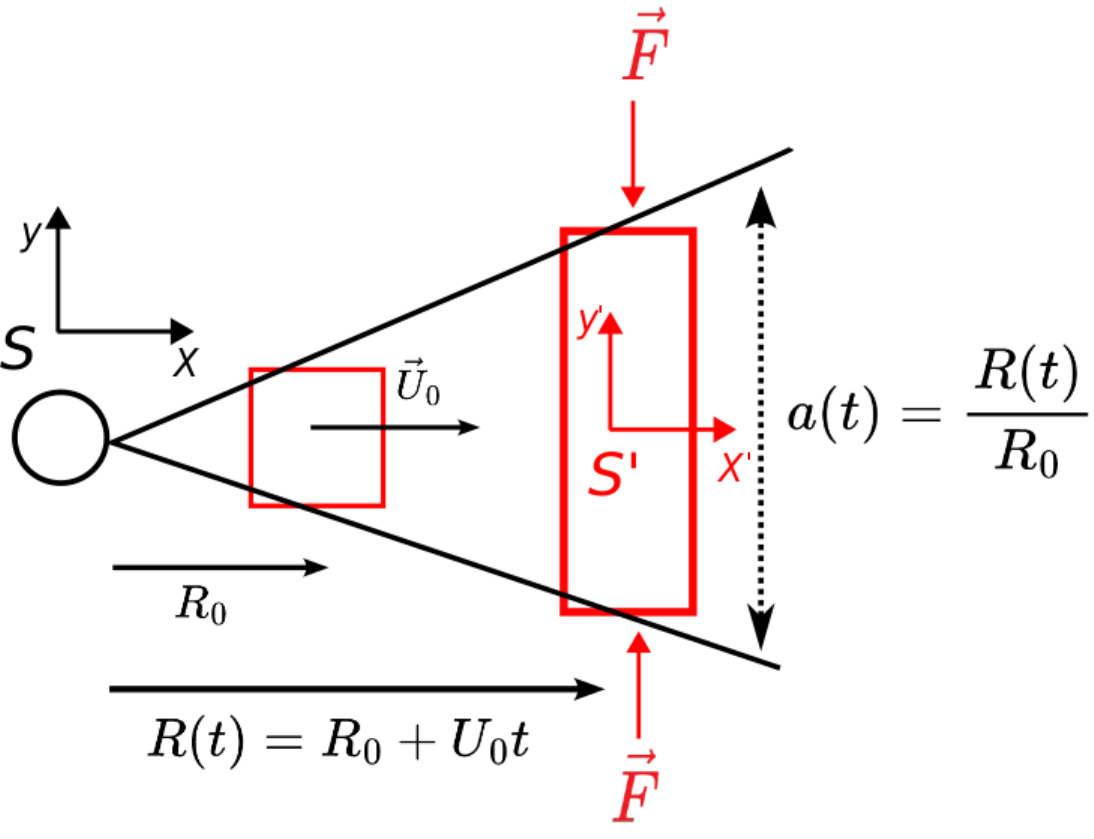}
    \caption{Expanding Box Model and the related quantities. The radial-spherical expansion of the plasma is approximated through a Cartesian description of coordinates. The box is expanding with constant velocity $\textbf{U}_0$ and the radial distance $R(t)$ is related to the Galilean transformation in the $x$ direction. In the perpendicular direction ($y$ and $z$ coordinates), respect to the expansion, the re-normalization yields to non-inertial fictitious forces $\textbf{F}$ that maintain a constant volume in the box.}
    \label{fig: EBM aproximation}
\end{figure}

The expanding box model (EBM) is a formalism that allows studying the radial-spherical expansion through a Cartesian approximation with a non-trivial change of coordinates. Consider a plasma parcel expanding from a static and inertial system of reference $S$; the EBM defines a new system of reference $S'$ co-moving with the plasma at constant velocity $\textbf{U}_0$. The plasma parcel is not expanding with respect to this new framework (through the change of coordinates),  but it moves together with it \cite[see][for a detailed discussion and derivation] {grappin.1993, Grappin.1996}. The relationship between these two systems, $S$ (non-prime quantities) and $S'$ (prime quantities), is defined as a Galilean transformation in the radial direction $x$ and a re-normalization in the perpendicular direction
\begin{align}
    x' &= x - R(t)\,,\label{e2.1}\\
    y' &= \frac{1}{a(t)} y\,,\label{e2.2}\\
    z' &= \frac{1}{a(t)} z \label{e2.3}\,,
\end{align}
where
\begin{align}
    R(t) &= R_0 + U_0 t\,,\label{e2.4}\\
   a(t) &\equiv \frac{R(t) }{ R_0} = 1 + \frac{U_0}{R_0}t\,.\label{e2.5}
\end{align}
The Galilean transformation in the $\hat{x}$ direction, given by Eq.~\eqref{e2.1}, allows the $S'$ system to move along with the plasma parcel as this transformation is made through the radial distance $R(t)$ of the plasma. In the transverse directions (i.e., $y$ and $z$ components) the re-normalization of the quantities is made through the expanding parameter $a(t)$. As $y$ and $z$ coordinates increase at the same rate as this parameter, the normalization \eqref{e2.2} and \eqref{e2.3} allows to maintain a constant volume of the plasma parcel in the $S'$ system or co-moving frame. It is important to stress the following interpretation for the coordinates transformations: note that only the radial direction is related to a Galilean transformation, but in the perpendicular directions there is a non-Galilean time-dependent renormalization in the quantities. These non-Galilean terms are expected to incorporate all the expanding effects in the kinetic equations. { Fig. \ref{fig: EBM aproximation} shows the Cartesian sketch and the related coordinates and parameters that allow us to work in the co-moving $S'$ system. With this transformation, all information about the plasma expanding in the perpendicular direction ($y-z$ plane) is incorporated and quantified by the $a(t)$ parameter. Even though in the co-moving system the plasma is not expanding, this parameter will allow us to include the effects of the expansion in the equations, by changing the time and spatial derivatives, among other physical quantities. }

In order to describe the physics in the $S'$ system, or co-moving framework, we need to transform all the physical quantities from $S$ to $S'$. For the detailed derivations, the interested reader can consult \cite{grappin.1993}, \cite{Grappin.1996}, \cite{Liewer.2001} and \cite{Moya.2012}. In this work, we follow the same procedures for a re-derivation of these quantities in the EBM. One particular difference in our description is that the spatial and velocity gradients in the $S'$ system are defined as
\begin{equation}
    \nabla' = \left( \frac{\partial}{\partial x'}, \frac{\partial}{\partial y'}, \frac{\partial}{\partial z'}  \right)\,,\quad
    \nabla_{\textbf{v}'} = \left(\frac{\partial}{\partial v_x'}, \frac{\partial }{\partial v_y'}, \frac{\partial }{\partial v_z'} \right)\,.
\end{equation}
These expressions will allow us to have all the expanding transformations through, i.e., the expanding parameter $a$, explicitly outside the gradient definitions. As a consequence of that, when studying the expansion dynamics analytically it becomes more clear where the explicit time dependencies are. The same consideration applies to all quantities, i.e. bulk velocity, pressure and heat fluxes.

To derive the equations, first we need to align the $\hat{x}$ axis with the spherical radial coordinate. Following the transformations \eqref{e2.1}-\eqref{e2.3}, we can establish more relationships between $S$ and $S'$ (see \ref{section: appendix A} for the detailed derivation)
\begin{align}
    \textbf{U}_0 &= U_0 \left( \hat{x} + \frac{y'}{R_0} \hat{y} + \frac{z'}{R_0}\hat{z}\right)\,, \label{e2.8}\\
    \textbf{v} &= \mathbb{A}\cdot \textbf{v}' + \textbf{U}_0 \,, \label{e2.9}\\
    \frac{\partial}{\partial t} &= \frac{\partial }{\partial t'} - \textbf{D}\cdot \nabla'\,, \label{e2.10}\\
    \nabla &=   \mathbb{A}^{-1} \cdot \nabla'\,, \label{e2.11}\\
    \nabla_\textbf{v} &=  \mathbb{A}^{-1}\cdot \nabla_{\textbf{v}'}\,, \label{e2.12}
\end{align}
where
\begin{equation*}
        \textbf{D} = U_0 \left(1, \frac{y'}{R}, \frac{z'}{R} \right)\,,
      \quad   \mathbb{A} (t)= \begin{pmatrix}
    1 & 0 & 0 \\
    0 & {a(t)} & 0 \\
    0 & 0 & {a(t)}
    \end{pmatrix}\,.
\end{equation*}
Note that $\textbf{v}' =  \left(v_x', v_y', v_z' \right)$ and $\nabla ' = \left(\frac{\partial}{ \partial x'}, \frac{\partial}{\partial y'},\frac{\partial}{\partial z'} \right)$, as discussed before. 
{These relationships are now sufficient to re-write Vlasov's equation in the co-moving system $S'$.}

\subsection{Motivation for an expanding Vlasov equation approach}

The effects of the expansion in the transverse directions (perpendicular to the radial direction $x$) can be directly observed and measured only from an inertial framework, such as the one fixed to the Sun. 
In the non-inertial framework $S'$ of the EBM, the effects of the expansion on the main plasma parameters are indirectly transmitted by the new parameter $a(t)$ that varies in time according to equation \eqref{e2.5}. This helps us to understand how we can study plasma expansion in the co-moving system. In fact, we convert the spatial and temporal  evolution of the expanding plasma properties to a temporal variation/description of the same properties in the $S'$ system. Therefore, even if the plasma parcel is not expanding in the co-moving frame, the expansion effects are transmitted to the equations via the $a (t)$ parameter. Moreover, the expanding box is not closed or isolated but should allow particle and energy exchange with the environment plasma.

Even though there are diverse applications of the EBM in, e.g., simulations at different scales, no kinetic theory applies this model from first principles. We need to start from the Vlasov equation in order to develop a fundamentally reliable framework for the study of plasma expansion in the EBM. This equation allows us to study the evolution of the velocity distribution function (VDF) for each particle population that compounds the plasma. Through the EBM formalism, we can study how expansion affects this equation. As mentioned, the expansion is mainly traduced through the directions perpendicular to the radial direction, and in the co-moving frame, there are non-inertial forces that maintain the constant volume of the plasma parcel. Thus, these forces will modify the acceleration term in Vlasov's equation. 

In the EBM, the modified expanding-Vlasov equation facilitates the multi-scale physical analysis of the expanding plasma.
For instance, the macroscopic properties of the expanding plasma are given by the main moments of the velocity distribution, while the equations describing their time-space evolution, i.e. continuity, momentum, and energy equations, are obtained by integrating the modified Vlasov equation. In this way, we are entitled to compare our results with those already published \cite{Grappin.1996, Liewer.2001}, but this time starting from the first principles in the derivation of the Vlasov equation. We recall that in the cited literature, the expanding MHD equations were obtained directly from transforming the physical quantities for the continuity, momentum, and pressure equations to the co-moving frame. The applications are multiple, and as a first choice, we can use it to explore linear and quasilinear properties of plasma waves and instabilities, expected to govern the dynamics of poorly collisional plasmas from space.

\section{Vlasov and Maxwell Equations}
\label{section 3}
\subsection{Vlasov Equation}

In this section, we transform the Vlasov equation into the non-inertial system $S'$. We start from the kinetic equation for a collisionless magnetized plasma in the inertial system $S$
\begin{align}
    \frac{\partial f_\mu}{\partial t} + \textbf{v} \cdot \frac{\partial f_\mu}{\partial \textbf{r}} + \textbf{a} \cdot \frac{\partial f_\mu}{\partial \textbf{v}} = 0\,,\label{eq: vlasov}
\end{align}
where $\mu$ represents the species that may be relevant for the plasma system, e.g., electrons, protons, and heavier ions. In order to obtain this equation in the new system $S'$, we must re-write every derivative and physical quantity. In addition, as in many astrophysical environments, magnetic field effects are dominant, for the electromagnetic field transformation we consider the so-called ``magnetic limit"; i.e. $E \sim (U_0/c) B \ll B$, with $c$ the speed of light \cite{rousseaux2013}. Namely, 
\begin{align}
    \textbf{E} = \textbf{E}' - \frac{1}{c} \textbf{U}_0 \times \textbf{B}'\,,\;\;\; \textbf{B} = \textbf{B}'\,. \label{e2.14}
\end{align}
As in this work we aim to describe non-relativistic plasmas, the transformations given by \eqref{e2.14} have been obtained in the low-velocity limit ($v/c \ll 1 $) and constitute a Galilean transformation (please note that Eq. \eqref{e2.14} follows assumptions that may require further elaboration in future works). A relativistic plasma description can be found in Ref. \cite{rousseaux2013}, where extensive discussions and formalism are described for the constitutive equations in both relativistic and Galilean limits, for electric and magnetic fields. Nevertheless, note that, in the EB context these transformations are not purely Galilean. Even though they have the same form as the Galilean's, we recall that the quantities are projected into the co-moving frame. For instance, the coordinates of the expanding velocity $\textbf{U}_0$ are written in terms of the EB variables, given by Eq. \eqref{e2.8}. Therefore, these transformations are not purely Galilean, and these terms are expected to give us the plasma's expanding properties. Eqs. \eqref{e2.14} have been widely studied and applied to different physical scenarios. In the EBM context, several applications are found in both theory and simulations; see, for example, references \cite{grappin.1993,Grappin.1996, Liewer.2001, Moya.2012} where they used the transformations given by Eq. \eqref{e2.14} to develop macroscopic physics in the EB frame.

On the other hand, and also non-relativistic conditions, the acceleration $\textbf{a}'$ (species of sort $\mu$) in the $S'$ system reads
\begin{align}
    \textbf{a}' = \frac{q_\mu}{m_\mu} \mathbb{A}\cdot \left( \textbf{E}' + \frac{1}{c} \left[ 
 \mathbb{A}\cdot \textbf{v}'\right]\times \textbf{B}'\right) - \frac{U_0}{R_0} \, \mathbb{T}\cdot \textbf{v}'\,, \label{e2.15}
\end{align}
where the first terms related to the electric and magnetic fields are the Lorentz force written in the co-moving frame, and $q_\mu$ and $m_\mu$ are the charge and mass of species of sort $\mu$. The last term represents the non-inertial result of the transformation, and $\mathbb{T}$ is a diagonal matrix that projects a vector in the perpendicular direction respective to the expansion
\begin{align}
    \mathbb{T} = \begin{pmatrix}
    0 & 0 & 0 \\
    0 & 1 & 0 \\
    0 & 0 & 1
    \end{pmatrix}\,. \label{e2.16}
\end{align}
Thus, with the definitions described above, the Vlasov equation in the co-moving frame read as 
\begin{align}
    \frac{\partial f'_\mu}{\partial t'} + \textbf{v}' \cdot \nabla' f'_\mu 
+ \frac{q_\mu}{m_\mu}\left\{ \textbf{E}' + \frac{1}{c} \left[\mathbb{A}\cdot \textbf{v}'\right]\times \textbf{B}' \right\} \cdot \left[ \nabla_{\textbf{v}'} f'_\mu \right] 
= \frac{U_0}{R_0} \left( \mathbb{T} \cdot \textbf{v}' \right) \cdot \left[\left(\mathbb{A}^{-1}\cdot \nabla_{\textbf{v}'}\right) f'_\mu \right]\,. \label{eq: Vlasov eq in S'}
\end{align}
See \ref{section: appendix A} for details of its derivation.

The modified Vlasov's equation \eqref{eq: Vlasov eq in S'} has explicit dependence on the expanding properties expressed in the EBM framework. Note that the left side of the equation keeps the same form as the one in the inertial frame ($S$ system), but with primed quantities. This is mainly because of the following two reasons: (1)~the first two terms in Eq.~\eqref{eq: vlasov} is a total derivative $df/dt$, as we are working in a non-relativistic frame $d t= d t'$, and the total time derivatives are invariant to the transformation from $S$ to $S'$; (2) The Lorentz force from the acceleration \eqref{e2.15} is multiplied by the matrix $\mathbb{A}$ that is compensated by its inverse $\mathbb{A}^{-1}$ in the inner product with velocity gradient of the distribution function. The only expanding modification in the Lorentz force is due to velocity. On the right-hand side of the Vlasov equation there is a new term relating to the non-inertial force from the second term in Eq.~\eqref{e2.15}. Despite the fact that in this model the co-moving frame is not accelerating, the re-normalization in the transverse components (i.e. $y'$ and $z'$) leads to a non-inertial force that maintains a constant volume in the box: work must be done upon the walls of the box to compensate the expansion of the plasma. This force contains all the expanding information related to the EBM, and now implicitly modifies Vlasov's equation.

In this paper, we aim to introduce such a framework for studying kinetic physics using the EBM. There are different possibilities and paths to follow. On one side, we can obtain fluid equations by integrating Eq.~\eqref{eq: Vlasov eq in S'} and obtaining the moments of the Vlasov equation, i.e., continuity, momentum, and pressure equations. 
This first step would allow us to compare the already published work with the results derived from the Vlasov equation. On the other hand, we could also study kinetic physics by coupling to the electromagnetic fields self-consistently described by the Maxwell equations. 
The advantage of having an extended description of the Vlasov equation is that we can now study expansion effects with both micro and macroscopic approaches. Aiming to provide an example of how we can work with the modified Vlasov equation, in the following sections we will derive the expanding moments of equation \eqref{eq: Vlasov eq in S'}, which further enables obtaining an ideal MHD description of the expanding plasma.

\subsection{Maxwell Equations}

Maxwell equations have been widely studied in the EB context, see for example \cite{Innocenti_2019}. In this section, we will only re-write those equations in terms of our notation in order to have the complete set of kinetic equations. Maxwell equations in the $S$ system read as
\begin{align*}
    \nabla \cdot \textbf{E} &= 4 \pi \rho\,, \qquad \nabla \cdot \textbf{B} = 0\,, \\
    \nabla \times \textbf{E} &= - \frac{1}{c} \frac{\partial \textbf{B}}{\partial t}\,, \;\;\; \nabla \times\textbf{B} = \frac{4 \pi}{c} \textbf{J} + \frac{1}{c} \frac{\partial \textbf{E}}{\partial t}\,.
\end{align*}

According to the transformations of the electromagnetic fields \eqref{e2.14}, Maxwell's equations in the co-moving frame read as
\begin{align}
     \left(\mathbb{A}^{-1} \cdot  \nabla'\right) \cdot \textbf{E}' &= 4 \pi a^2  \rho' + \frac{1}{c} \left(\mathbb{A}^{-1} \cdot  \nabla'\right) \cdot \left[ \textbf{U}_0 \times \textbf{B}' \right]\,, \label{e17}\\  
     \left(\mathbb{A}^{-1}\cdot \nabla' \right) \cdot \textbf{B}' &= 0\,,\\
     \left( \mathbb{A}^{-1}\cdot \nabla'\right)\times\textbf{E}' &= -\frac{1}{c}\frac{\partial \textbf{B}'}{\partial t'} - \frac{U_0}{R c}\mathbb{L}\cdot \textbf{B}' \,, \\
     \left(\mathbb{A}^{-1} \cdot  \nabla'\right)  \times \textbf{B}' &= \frac{4 \pi}{c}\left[ a^2 \mathbb{A}\cdot\textbf{J}' + a^2 \rho' \textbf{U}_0\right] + \frac{1}{c} \frac{\partial\textbf{E}'}{\partial t'} +  \frac{1}{c^2} \left[\textbf{U}_0 \cdot\left(\mathbb{A}^{-1}\cdot \nabla'\right)\right] \left( \textbf{U}_0 \times \textbf{B}'\right) \nonumber \\ & - \frac{1}{c}\left[\mathbf{U}_0 \cdot \left(\mathbb{A}^{-1} \cdot \nabla'\right)\right] \textbf{E}'  - \frac{1}{c^2} \textbf{U}_0 \times \frac{\partial \textbf{B}'}{\partial t'}    \label{e20}\,,
\end{align}
where the charge density and current between both systems relate according to
\begin{align}
    \rho &= a^2 \rho'\,, \label{e21}\\
    \textbf{J} &= a^2 \mathbb{A}\cdot\textbf{J}' + a^2 \rho' \textbf{U}_0\,, \label{e22}
\end{align}
and
\begin{align}
\rho ' &= q \int f'\left(\textbf{r}',\textbf{v}',t '\right) d \textbf{v} ' \,,\\
\textbf{J}' &= q \int \textbf{v}' f'\left(\textbf{r}',\textbf{v}',t' \right) d \textbf{v}'\,,\\
    \mathbb{L} &= \begin{pmatrix}
         2 & 0 & 0 \\
    0 & 1 & 0 \\
    0 & 0 & 1
    \end{pmatrix}\,.
\end{align}

The $a^2$ parameter in Eqs. \eqref{e21} and \eqref{e22} explicitly appears when transforming the velocity differential between both system through the Jacobian matrix as $d \textbf{v} = a^2 d \textbf{v}'$ in the density and current integrals. Note that these Equations can be simplified considering non-relativistic plasma $ U_0/c \ll 1$. Nevertheless, for consistency, all the terms of the equations are shown. It is worth mentioning that Maxwell's equation can be simplified for the magnetic limit in the transformations of the field \eqref{e2.14}. In such cases, where the magnetic effects govern plasma dynamics, the displacement current can be neglected in Ampere's Law, which reduces significantly the complexity of these equations (see \cite{rousseaux2013} for Galilean Electromagnetism). In the co-moving frame and for the magnetic limit, Maxwell's equations read as
\begin{align}
    \left(\mathbb{A}^{-1} \cdot  \nabla'\right) \cdot \textbf{E}' &= 4 \pi a^2  \rho' + \frac{1}{c} \left(\mathbb{A}^{-1} \cdot  \nabla'\right) \cdot \left[ \textbf{U}_0 \times \textbf{B}' \right]\,, \label{e26.1}\\  
     \left(\mathbb{A}^{-1}\cdot \nabla' \right) \cdot \textbf{B}' &= 0\,,\\
     \left( \mathbb{A}^{-1}\cdot \nabla'\right)\times\textbf{E}' &= -\frac{1}{c}\frac{\partial \textbf{B}'}{\partial t'} - \frac{U_0}{R c}\mathbb{L}\cdot \textbf{B}' \,, \\
     \left(\mathbb{A}^{-1} \cdot  \nabla'\right)  \times \textbf{B}' &= \frac{4 \pi}{c}\left[ a^2 \mathbb{A}\cdot\textbf{J}' + a^2 \rho' \textbf{U}_0\right] \,. \label{e29}
\end{align}
We recall that both sets of equations are written in the EB frame; the First one\eqref{e17}-\eqref{e20} explicitly shows how the displacement current changes between both frames. On the other hand, in the magnetic limit this current can be neglected and Maxwell's equations are described by Eqs. \eqref{e26.1} - \eqref{e29}.

\section{Moments of Vlasov's Equation}
\label{section 4}

In this section, we will focus on developing equations for a magnetized fluid with the EB formalism. This first approach is a possible example of working with Vlasov's equation in the co-moving frame. To do so, we first need to obtain the first three moments from Eq.~(\ref{eq: Vlasov eq in S'}): continuity, momentum and pressure. We stress that these moments are expressed in the non-inertial co-moving frame $S'$. As Eq.~\eqref{eq: Vlasov eq in S'} is now written in the EB frame, we only need to integrate this equation by the primed moments in the velocity space (similar to the non-expanding cases).
  
Note that the left-hand side of Eq.~\eqref{eq: Vlasov eq in S'} is almost the same as in the non-expanding case. Therefore, the same ideas used in that case are still valid but with primed quantities. Note that the modification of the equation is mainly through the tensor $\mathbb{A}(t)$. One advantage of this description is that moments are obtained via integrating Vlasov's equation in the velocity space. As this tensor is only a function of time, such integrals will not be affected (only the components through the inner products). Therefore, we will only focus on the right-hand side integrals of Eq.~\eqref{eq: Vlasov eq in S'}. For the notation and solution of the related integrals, see \ref{section: appendix B}.
 
For developing the equations we will follow the ideas of \textcite{Hunana_2019}. Related to the moments of the velocity distribution function $f'$ (density, mean velocity, pressure and heath flux), we will use the same definitions as in the non-expanding cases, but with primed quantities
\begin{align}
    n' &= \int f'(\textbf{r}', \textbf{v}', t') d \textbf{v}' \,. \label{e3.1}\\
    \textbf{u}' &= \frac{1}{n'} \int \textbf{v}' f'(\textbf{r}', \textbf{v}', t') d \textbf{v}'\,, \label{e3.2}\\
    \mathbb{P}' &= m \int \textbf{w}' \textbf{w}' f'(\textbf{r}', \textbf{v}', t') d \textbf{v}' \,, \label{e3.3}\\
    \mathbb{Q}' &= m \int \textbf{w}' \textbf{w}' \textbf{w}' f'(\textbf{r}', \textbf{v}', t') d \textbf{v}'\,, \label{e3.4}
\end{align}
where the velocity $\textbf{w}'$ denotes the fluctuations with respect to the mean velocity $\textbf{u}'$, so that
\begin{align*}
    \textbf{v}' = \textbf{u}' + \textbf{w}'\,.
\end{align*}
Since we will continue to work only in the co-moving system, the primes in the variable notations will be omitted in the equations from now on.

\subsection{First Moment: Continuity Equation}

As a first example of working out the equations in the $S'$ system, we will develop in detail the first moment. For the other moments, the same ideas are applied. When we integrate Eq.~(\ref{eq: Vlasov eq in S'}) in the velocity space, from the first two terms we have the usual expression from the continuity equation (omitting the primes):
\begin{align*}
    \int \left( \frac{\partial f}{\partial t} + \textbf{v} \cdot \nabla f  \right) d \textbf{v} = \frac{\partial n}{\partial t} + \nabla \cdot \left(n \textbf{u} \right)\,.
\end{align*}

The Lorentz force terms are still zero as for the non-expanding case. For example, for the electric field
\begin{align*}
    \int \textbf{E} \cdot \left[   \nabla_{\textbf{v}} f \right] d \textbf{v} = E_i  \int \frac{\partial f}{\partial v_i} d \textbf{v} = 0\,,
\end{align*}
coupled to the condition
\begin{align*}
    \lim_{v \to \pm \infty } f (v) = 0\,.
\end{align*}

Finally, the integral reads as
\begin{align*}
    I & = \frac{U_0}{R_0} \int  \left( \mathbb{T} \cdot \textbf{v} \right) \cdot \left[\left(\mathbb{A}^{-1}\cdot \nabla_{\textbf{v}}\right) f \right] d \textbf{v}\\
    & = \frac{U_0}{R_0} \int \left( \frac{v_y }{a} \frac{\partial f}{\partial v_{y}} +  \frac{v_z}{a}  \frac{\partial f}{\partial v_{z}}  \right) d \textbf{v}\,.
\end{align*}
The two terms in the integral are similar, hence, basically, we are solving the same integral twice. Let us fix one coordinate (say the $y$-coordinate) and multiply that expression by 2 and integrate by parts, which yields
\begin{align*}
    I &= \frac{2 U_0}{a R_0} \int v_y \frac{\partial f}{\partial v_y} d \textbf{v}
     =  - \frac{2 U_0}{a R_0}  \int f d \textbf{v} = - \frac{2 U_0}{a R_0} n \,.
\end{align*}

Therefore, the continuity equation reads as follows in the co-moving system 
\begin{align}
    \frac{\partial n}{\partial t} + \nabla\cdot \left( n \textbf{u}\right) = - \frac{2 U_0}{ a R_0} n\,.
    \label{eq: continuity eq EBM}
\end{align}

With this first example, we can already start noticing how the expansion affects the equations through the expanding parameter $a(t)$. The continuity equation~\eqref{eq: continuity eq EBM} has a non-zero term on the right-hand side that modifies the behavior of the density. As we mentioned before, even though in the co-moving frame, the box is not spatially expanding, all the expanding information is traduced in time variations by the $a(t)$ parameter, allowing us to study how the expansion affects the physical quantities (i.e., density, velocity, and pressure) we are interested in. It is essential to mention that the obtained equation~\eqref{eq: continuity eq EBM} is the same as the one published by \textcite{grappin.1993}. Here, we are presenting the kinetic formalism for deducing the fluid equations.

\subsection{Second and Third Moment: Momentum and Pressure Equation}

For the second and third moment, we multiply Vlasov's equation by $\textbf{v}$ and $\textbf{v} \textbf{v}$, respectively, and integrate it over velocity space in order to obtain the momentum and pressure equation. For the detailed derivation see \ref{section: appendix B}. For these cases we will only focus in the right-hand side integrals related to the non-inertial force. The integral $\textbf{I}$ for the momentum equation reads as
\begin{align}
    \textbf{I}_j & = \frac{U_0}{R_0} \int \left\{\left( \mathbb{T} \cdot \textbf{v} \right) \cdot \left[\left(\mathbb{A}^{-1}\cdot \nabla_{\textbf{v}}\right) f \right] \textbf{v}\right\}_j d \textbf{v} \nonumber
     \\
     &=\frac{U_0}{a R_0}\int T_{ii} v_i  \frac{\partial f}{\partial v_i} v_j d \textbf{v}\nonumber\\
     & = - \frac{U_0}{a R_0} \int \left( T_{ii} v_j + T_{ii} v_i \right) f d \textbf{v} \nonumber \\
     & = - \frac{n U_0}{a R_0} \left\{\left[ 2 \mathbb{I} + \mathbb{T} \right]\cdot \textbf{u}\right\}_j\,.
\end{align}
Therefore, the expanding momentum equation becomes
\begin{align}
    \frac{\partial \textbf{u}}{\partial t}  + \left( \textbf{u} \cdot  \nabla\right) \textbf{u} = \frac{q}{m} \left[ \textbf{E} + \frac{1}{c} \left( \mathbb{A} \cdot \textbf{u} \right) \times \textbf{B} \right] 
    - \frac{1}{\rho} \nabla \cdot \mathbb{P}  - \frac{U_0}{a R_0} \mathbb{T}\cdot \textbf{u}\,.\label{eq: reduced momentum eq EBM} 
\end{align}

On the other hand, for the third moment/pressure equation the right-hand side integral $\mathbb{I}$ reads as
\begin{align*}
    \mathbb{I}_{jk} & = \frac{U_0}{R_0} \int\left\{ \left( \mathbb{T} \cdot \textbf{v}' \right) \cdot \left[\left(\mathbb{A}^{-1}\cdot \nabla_{\textbf{v}'}\right) f \right] \textbf{v} \textbf{v} \right\}_{jk} d \textbf{v} \\
   & = \frac{U_0}{a R_0} \int T_{ii} v_i  \frac{\partial f}{\partial v_i} v_j v_k d \textbf{v}\\
    & = - \frac{U_0}{a R_0} \left( 2 \int v_j v_k f d \textbf{v} +  T_{jj} \int v_j v_k f d \textbf{v} + T_{kk} \int v_k v_j f d \textbf{v}\right)\,.
\end{align*}
For solving these integrals, we only need to decompose the velocity in terms of the mean velocity and its fluctuations $\textbf{v} = \textbf{u} + \textbf{w}$, obtaining 
\begin{align}
    \mathbb{I} = - \frac{ 2 U_0}{a R_0} \left(\textbf{u} \textbf{u} n + \frac{1}{m} \mathbb{P} \right)- \frac{U_0}{a R_0} \left[  \mathbb{T}\cdot \left( \textbf{u}\textbf{u} n + \frac{1}{m}\mathbb{P} \right)  \right]^s,
\end{align}
where the supra-index $s$ represents a symmetric operator that acts on a matrix $\mathbb{C}$ as $\mathbb{C}^s = \mathbb{C} + \mathbb{C}^{T}$ or in index notation $C_{ij} = C_{ij} + C_{ji}$.
Finally, using the previous moments \eqref{eq: continuity eq EBM} and \eqref{eq: reduced momentum eq EBM}, the pressure equation takes the following form:
\begin{align}
    \frac{\partial \mathbb{P}}{\partial t} + \nabla \cdot \left( \textbf{u} \mathbb{P} + \mathbb{Q} \right) + \left[ \mathbb{P} \cdot \nabla \textbf{u} \right]^s + \frac{q}{m c} \left[  \textbf{B} \times 
 \left( \mathbb{A} \cdot \mathbb{P} \right) \right]^s 
    = - \frac{2 U_0}{a R_0} \mathbb{P} - \frac{U_0}{a R_0} \left[ \mathbb{T} \cdot \mathbb{P}\right]^s\,, \label{eq: reduced pressure eq EBM}
\end{align}
where the cross product between a vector $\textbf{V}$ and a tensor $\mathbb{C}$ is defined as
\begin{align*}
    \left[ \textbf{V} \times \mathbb{C} \right]_{ik} = \epsilon_{ijl}V_j C_{lk}\,.
\end{align*}

The obtained moments \eqref{eq: continuity eq EBM}, \eqref{eq: reduced momentum eq EBM} and \eqref{eq: reduced pressure eq EBM} are completely general regarding the Expanding Box Model. The expansion effects are clearly visible and change the usual (non-expanding) equations. In particular, the right-hand sides of the equations are now affected by the EBM terms (i.e., the $a$ parameter, the expanding velocity $U_0$ and the initial distance $R_0$). For the cases of the continuity and momentum equation, the terms $- \frac{2 U_0}{ a R_0} n$ and $ - \frac{U_0}{a R_0} \mathbb{T}\cdot \textbf{u}$, are the same as \textcite{grappin.1993} developed from transforming the continuity and momentum equation from the $S$ to the $S'$ system/co-moving frame. Here, we developed another framework to obtain these equations. The advantage of working from Vlasov's equation with the EBM is that we can explore the kinetic effects and compare our results with the already published work in MHD. 

On the other hand, it is expected that if the plasma is not expanding, we should recover the usual magnetized fluid equations. Note that this case is obtained in the non-expanding limit $U_0 = 0$ or $a = 1$, therefore $\mathbb{A}_{ij} = \delta_{ij}$ is the identity, and we recover the usual expressions for the moments of Vlasov's equation in the inertial frame, namely
\begin{align}
    &\frac{\partial n}{\partial t} + \nabla\cdot \left( n \textbf{u}\right) = 0\,, \label{e26}\\
    &\frac{\partial \textbf{u}}{\partial t}  + \left( \textbf{u} \cdot  \nabla\right) \textbf{u} = \frac{q}{m} \left[ \textbf{E} + \frac{1}{c}  \textbf{u}  \times \textbf{B} \right]  - \frac{1}{\rho} \nabla \cdot \mathbb{P}\,, \label{e27}\\
    &\frac{\partial \mathbb{P}}{\partial t} + \nabla \cdot \left( \textbf{u} \mathbb{P} + \mathbb{Q} \right) + \left[ \mathbb{P} \cdot \nabla \textbf{u} \right]^s + \frac{q}{m c} \left[ \textbf{B} \times 
   \mathbb{P}  \right]^s = 0 \,.\label{e28}
\end{align}

In the next section, we will focus on obtaining the ideal MHD equations with the EBM formalism in order to compare with the work of \textcite{grappin.1993} and prove the consistency of the obtained equations. As an improvement of the published equations, in this work we are also presenting the evolution for the pressure tensor given by Eq.~\eqref{eq: reduced pressure eq EBM}. This allows us to study how the expansion also affects the pressure tensor.

\section{Expanding Ideal-MHD}
\label{section 5}

In order to compare our results with the ones already published in the literature, we will develop, from the moments we already obtained, an ideal description for the magnetohydrodynamic equations using the expanding box model. In this section, we will explicitly work with the primed quantities so it is clear in which frame we are working.

Consider a magnetized electron-proton plasma, where the momentum equation (\ref{eq: reduced momentum eq EBM}) in the co-moving frame describes the  evolution for each species
\begin{align}
      \frac{\partial \textbf{u}_p'}{\partial t'}  + \left( \textbf{u}_p' \cdot  \nabla'\right) \textbf{u}_p' &= \frac{q_p}{m_p} \left[ \textbf{E}' + \frac{1}{c} \left( \mathbb{A} \cdot \textbf{u}_p' \right) \times \textbf{B}' \right] 
    - \frac{1}{\rho_p'} \nabla' \cdot \mathbb{P}_p'  - \frac{U_0}{a R_0} \mathbb{T}\cdot \textbf{u}_p'\,, \label{e4.1}\\
      \frac{\partial \textbf{u}_e'}{\partial t'}  + \left( \textbf{u}_e' \cdot  \nabla'\right) \textbf{u}_e' &= \frac{q_e}{m_e} \left[ \textbf{E}' + \frac{1}{c} \left( \mathbb{A} \cdot \textbf{u}_e' \right) \times \textbf{B}' \right] 
    - \frac{1}{\rho_e'} \nabla' \cdot \mathbb{P}_e'  - \frac{U_0}{a R_0} \mathbb{T}\cdot \textbf{u}_e'\,. \label{e4.2}
\end{align}
Defining the fluid velocity as 
\begin{align}
    \textbf{u} = \frac{m_e \textbf{u}_e + m_p \textbf{u}_p}{m_e + m_p} \approx \textbf{u}_p\,,
\end{align}
for $\frac{m_e}{m_p} \ll 1$ in Eq.~\eqref{e4.2} and in the cold electron approximation, ($\mathbb{P}_e' = 0$) we can relate the magnetic and electric fields as
\begin{align}
    \textbf{E}' = - \frac{1}{c} \left(\mathbb{A} \cdot \textbf{u}_e' \right) \times \textbf{B}'\,.
    \label{eq: electric and magnetic field related with the electrons velocity}
\end{align}

In the inertial system, assuming quasi-neutrality ($n_e = n_p = n$), for the the total current density $\textbf{J}$ we obtain 
\begin{align}
    \textbf{J} = q_p n \textbf{u}_p + q_e n \textbf{u}_e = e n \left(\textbf{u}_p - \textbf{u}_e \right)\,,
\end{align}
where $e$ is the electron charge. Therefore, expressing $\textbf{J}$ in terms of the co-moving variables
\begin{align}
    \textbf{J} = \mathbb{A} \cdot \textbf{J}' = e n' \mathbb{A} \cdot \left(\textbf{u}_p' - \textbf{u}_e' \right),
\end{align}
where $\textbf{J}' \equiv  e n' \left(\textbf{u}_p' - \textbf{u}_e' \right)$, the electron velocity can be expressed as
\begin{align}
    \textbf{u}_e' = \textbf{u}_p' - \frac{1}{e n}   \textbf{J}' \,. \label{eq: electron's velocity in S'}
\end{align}
Therefore, using equation (\ref{eq: electron's velocity in S'}) in (\ref{eq: electric and magnetic field related with the electrons velocity}) we obtain
\begin{align}
    \textbf{E}' = \frac{1}{e n' c}\left(\mathbb{A} \cdot \textbf{J}'\right)\times \textbf{B}'  - \frac{1}{c} \left(\mathbb{A}\cdot \textbf{u}_p' \right) \times \textbf{B}' 
\end{align}

In the inertial frame, we can obtain the current density in terms of the magnetic field from Ampere's law. Neglecting the displacement current (for studying lower frequency waves as in the proton's scale), Ampere's law reads as
\begin{align}
    \nabla \times \textbf{B} = \frac{4 \pi}{c} \textbf{J}\,.
\end{align}
Therefore, in the co-moving system we have
\begin{align}
    \mathbb{A} \cdot \textbf{J}' = \frac{c}{4 \pi}\left(\mathbb{A}^{-1} \cdot \nabla '\right) \times \textbf{B}'\,.
\end{align}
Finally, the electric field reads as
\begin{align}
    \textbf{E}' = \frac{1}{4 \pi e n'} \left[\left(\mathbb{A}^{-1}\cdot \nabla' \right)\times \textbf{B}'\right] \times \textbf{B}' - \frac{1}{c} \left(\mathbb{A}\cdot \textbf{u}_p' \right) \times \textbf{B}' \,. \label{eq: electric field MHD}
\end{align}

The first term of this equation is related to the Hall term and the second term is the usual one from the MHD induction equation.
Finally, using Eq.~(\ref{eq: electric field MHD}) in the momentum equation for protons \eqref{e4.2}, we get
\begin{align}
    &\frac{\partial \textbf{u}_p'}{\partial t'} + \left(\textbf{u}_p' \cdot \nabla' \right)\textbf{u}_p' + \frac{1}{\rho_p} \nabla' \cdot \mathbb{P}_p'
    + \frac{1}{8 \pi \rho_p'} \left[\left(\mathbb{A}^{-1} \cdot \nabla' \right)\right] B'^2 \nonumber \\
    &- \frac{1}{4 \pi \rho_p'}  \left[\textbf{B}'\cdot \left(\mathbb{A}^{-1}\cdot \nabla' \right) \right] \textbf{B}'  = - \frac{U_0}{a R_0}\mathbb{T} \cdot \textbf{u}_p'\,, \label{eq: MHD equation for proton's velocity}
\end{align}
where $\rho_p'$ is the proton density. Equation~\eqref{eq: MHD equation for proton's velocity} is the same as the one obtained by \textcite{grappin.1993}, the only difference here is that we explicitly work with the $\mathbb{A}$ tensor, outside the definition of the gradient.

\subsection{Pressure Equation}

As a final application of the presented framework we will develop the polytropic equation from Eq.~\eqref{eq: reduced pressure eq EBM} that also was published in \textcite{grappin.1993, Grappin.1996}. For this case, consider an isotropic pressure $p'$ given by
\begin{align}
    \mathbb{P}' = \begin{pmatrix}
    p' & 0 & 0 \\
    0 & p' & 0 \\
    0 & 0 & p'
    \end{pmatrix}\,. \label{e40}
\end{align}
Neglecting the heat flux from \eqref{eq: reduced pressure eq EBM}, we obtain
\begin{align}
     \frac{\partial \mathbb{P}'}{\partial t'} + \nabla' \cdot \left( \textbf{u}' \mathbb{P}' \right) + \left[ \mathbb{P}' \cdot \nabla' \textbf{u}' \right]^s + \frac{q}{m c} \left[  \textbf{B}' \times 
 \left( \mathbb{A} \cdot \mathbb{P}' \right) \right]^s  
    = - \frac{2 U_0}{a R_0} \mathbb{P}' - \frac{U_0}{a R_0} \left[ \mathbb{T} \cdot \mathbb{P}'\right]^s \label{e41.1}
\end{align}
Calculating the trace, assuming a pressure tensor given by Eq. \eqref{e40}, we obtain the scalar pressure equation for a polytropic index $\gamma = 5/3$:
\begin{align}
    \frac{\partial p'}{\partial t'} + \textbf{u}'\cdot \nabla' p' + \gamma p' \nabla' \cdot \textbf{u}' = - \gamma \frac{2 U_0}{a R_0} p'\,. \label{e41}
\end{align}
This equation is the same as the one published in the cited works. In this section, we aimed to study the consistency between the micro and macroscopic approaches. The results show that we are able to reproduce expanding (-ideal) MHD through the modified Vlasov equation \eqref{eq: Vlasov eq in S'}. Developing an ideal MHD description allows us to validate the expanding Vlasov equation \eqref{eq: Vlasov eq in S'}. In order to study kinetic physics, we first needed to validate the kinetic equation. As there is already published work in the EBM-MHD frame, this is the natural starting point to apply the kinetic-EBM.  

\section{Equations Summary}

After obtaining all equations for different levels of plasma descriptions, in this section, we summarize our main results in the co-moving frame. Table \ref{table: summary} shows all obtained equations. The expanding (ideal-)MHD equations, which describe the plasma dynamics in the EBM frame, are given by the continuity equation \eqref{eq: continuity eq EBM}, the momentum equation for protons \eqref{eq: MHD equation for proton's velocity}, and the polytropic equation \eqref{e41}. The magnetic field equation is obtained by neglecting the Hall term in equation \eqref{eq: electric field MHD} when replacing it in Faraday's Law. 

\begin{table}[H]
\caption{Equation summary in the EB frame.}
    \centering
    \begin{tabular}{c|c}
         & Expanding/co-moving frame\\
         \hline \\
       \textbf{First Principles:} \\ Vlasov Equation  & $  \frac{\partial f'_\mu}{\partial t'} + \textbf{v}' \cdot \nabla' f'_\mu 
+ \frac{q_\mu}{m_\mu}\left\{ \textbf{E}' + \frac{1}{c} \left[\mathbb{A}\cdot \textbf{v}'\right]\times \textbf{B}' \right\} \cdot \left[ \nabla_{\textbf{v}'} f'_\mu \right] =
\frac{U_0}{R_0} \left( \mathbb{T} \cdot \textbf{v}' \right) \cdot \left[\left(\mathbb{A}^{-1}\cdot \nabla_{\textbf{v}'}\right) f'_\mu \right]$\,,  \\
         &\\
         &  $\left(\mathbb{A}^{-1} \cdot  \nabla'\right) \cdot \textbf{E}' = 4 \pi a^2  \rho' + \frac{1}{c} \left(\mathbb{A}^{-1} \cdot  \nabla'\right) \cdot \left[ \textbf{U}_0 \times \textbf{B}' \right]\,,$ \\
      Gauss' Law  &\\
         &  $\left(\mathbb{A}^{-1}\cdot \nabla' \right) \cdot \textbf{B}' = 0\,,$\\
         &\\
      Faraday's Law   & $ \left( \mathbb{A}^{-1}\cdot \nabla'\right)\times\textbf{E}' = -\frac{1}{c}\frac{\partial \textbf{B}'}{\partial t'} - \frac{U_0}{R c}\mathbb{L}\cdot \textbf{B}', $ \\
         & \\
         Ampere's Law & $ \left(\mathbb{A}^{-1} \cdot  \nabla'\right)  \times \textbf{B}' = \frac{4 \pi}{c} \left[ a^2 \mathbb{A}\cdot\textbf{J}' + a^2 \rho' \textbf{U}_0\right] + \frac{1}{c} \frac{\partial\textbf{E}'}{\partial t'} +  \frac{1}{c^2} \left[\textbf{U}_0 \cdot\left(\mathbb{A}^{-1}\cdot \nabla'\right)\right] \left( \textbf{U}_0 \times \textbf{B}'\right) \nonumber  $ \\
         &  \;\;\;$- \frac{1}{c}\left[\mathbf{U}_0 \cdot \left(\mathbb{A}^{-1} \cdot \nabla'\right)\right] \textbf{E}'  - \frac{1}{c^2} \textbf{U}_0 \times \frac{\partial \textbf{B}'}{\partial t'}  \,.$\\
         & \\
         \hline\\
        \textbf{Multi-Fluids:}  & \\
        Continuity Equation & $\frac{\partial n}{\partial t} + \nabla\cdot \left( n \textbf{u}\right) = - \frac{2 U_0}{ a R_0} n\,,$ \\
        & \\
        Momentum Equation & $\frac{\partial \textbf{u}}{\partial t}  + \left( \textbf{u} \cdot  \nabla\right) \textbf{u} = \frac{q}{m} \left[ \textbf{E} + \frac{1}{c} \left( \mathbb{A} \cdot \textbf{u} \right) \times \textbf{B} \right] 
    - \frac{1}{\rho} \nabla \cdot \mathbb{P}  - \frac{U_0}{a R_0} \mathbb{T}\cdot \textbf{u}\,,$ \\
        &\\
       Energy/Pressure Equation & $ \frac{\partial \mathbb{P}}{\partial t} + \nabla \cdot \left( \textbf{u} \mathbb{P} + \mathbb{Q} \right) + \left[ \mathbb{P} \cdot \nabla \textbf{u} \right]^s + \frac{q}{m c} \left[  \textbf{B} \times 
 \left( \mathbb{A} \cdot \mathbb{P} \right) \right]^s \nonumber 
    = - \frac{2 U_0}{a R_0} \mathbb{P} - \frac{U_0}{a R_0} \left[ \mathbb{T} \cdot \mathbb{P}\right]^s\,, $\\
    &\\
    \hline \\
    \textbf{Expanding (ideal-) MHD:} & \\
    Continuity Equation & $\frac{\partial n}{\partial t} + \nabla\cdot \left( n \textbf{u}\right) = - \frac{2 U_0}{ a R_0} n\,,$  \\
    & \\
    Momentum Equation & $\frac{\partial \textbf{u}}{\partial t} + \left(\textbf{u} \cdot \nabla \right)\textbf{u} + \frac{1}{\rho} \nabla p
    + \frac{1}{8 \pi \rho} \left[\left(\mathbb{A}^{-1} \cdot \nabla \right)\right] B^2 
    - \frac{1}{4 \pi \rho}  \left[\textbf{B}\cdot \left(\mathbb{A}^{-1}\cdot \nabla \right) \right] \textbf{B}  = - \frac{U_0}{a R_0}\mathbb{T} \cdot \textbf{u}\,,$ \\
    & \\
    Polytropic Equation & $ \frac{\partial p}{\partial t} + \textbf{u}\cdot \nabla p + \gamma p \nabla \cdot \textbf{u} = - \gamma \frac{2 U_0}{a R_0} p\,,$\\
    & \\
    Magnetic Field Equation & $\frac{\partial \textbf{B}}{\partial t} + \left(  \nabla \cdot \textbf{u} \right) \textbf{B} + \left(  \textbf{u} \cdot  \nabla \right) \textbf{B} 
   - \left[ \textbf{B}\cdot \left(\mathbb{A}^{-1} \cdot \nabla \right) \right]  \left( \mathbb{A}\cdot \textbf{u}\right) = - \frac{U_0}{a R_0} \mathbb{L} \cdot \textbf{B}\,, $\\
   & \\
   \hline 
    \end{tabular}
    \label{table: summary}
\end{table}

Where $a = a(t) = 1+U_0\,t/R_0$\,, and
\begin{align}
    \mathbb{L} = \begin{pmatrix}
         2 & 0 & 0 \\
    0 & 1 & 0 \\
    0 & 0 & 1
    \end{pmatrix}
    \,,\quad
     \mathbb{T} = \begin{pmatrix}
    0 & 0 & 0 \\
    0 & 1 & 0 \\
    0 & 0 & 1
    \end{pmatrix}
    \,,\quad 
     \mathbb{A}= \begin{pmatrix}
    1 & 0 & 0 \\
    0 & {a} & 0 \\
    0 & 0 & {a}
    \end{pmatrix}\,.
\end{align}

Equations in the last section of Table \ref{table: summary} describe the ideal MHD equations in the co-moving/EB frame. The right-hand side of the equations is no longer equal to zero (compared with the ideal MHD in the inertial frame), and there is a clear dependence on the expanding parameters. These modifications are related to the work done in the transverse directions ($y-$ and $z-$coordinates) of the box to maintain a constant volume in the new frame. For instance, the momentum equation clearly shows these non-inertial forces represented through the $\mathbb{T}$ tensor, which projects the velocity in the transverse directions.

\section{Discussions and Conclusions}

In this paper, we have developed a novel first principles Vlasov-based approach to describe astrophysical plasma expansion, using the Expanding Box Model (EBM) formalism. Under appropriate coordinate transformations, we transferred the analysis to a non-inertial frame co-moving with the expanding plasma parcel identified with the EBM. This EBM frame is characterized by a re-normalization of components that maintains a constant volume of the plasma parcel. Within this description, in the EBM frame, plasma does not expand, but its properties vary in time as an effect of the expansion. Thus, the expansion was traduced and simplified from complex spatio-temporal variations in the inertial frame (fixed to, e.g., the Sun), to purely local temporal variations in the co-moving non-inertial EBM frame, mainly represented by the bulk speed of the plasma $U_0$ and the $a(t)$ parameter. 

This procedure allowed us, for the first time, to rewrite all fundamental plasma physics equations with explicit dependence on the EBM parameters. In particular, through the coordinate transformation given by Eqs.~\eqref{e2.1} - \eqref{e2.3}, we derived the Vlasov equation in the EBM frame, which explicitly considers the non-inertial fictitious forces related to the expanding parameters. These coordinate transformations are not purely Galilean. In the transverse directions with respect to the expansion, there is a non-Galilean re-normalization of the quantities. From these transformations, it is possible to transform all the physical quantities (i.e., electric and magnetic fields, spatial and temporal derivatives, etc.) into the co-moving frame through a Galilean-like transformation, in which the variables are projected into the EB frame adding the non-Galilean/inertial terms. Therefore, the EBM formalism provides a new framework to study the kinetic and fluid dynamics of expanding plasma, to be applied to different expanding systems. For instance, the modified Vlasov equation allows us to study the expansion effects in the evolution of velocity distribution functions of plasma particles. We can couple it with Maxwell's equations and resolve kinetic spectra of plasma wave fluctuations, spontaneous or induced by the kinetic anisotropy of charged particles, their linear dispersion relations, quasi-linear and nonlinear interactions with plasma particles, etc.

To explore the physical interpretations of the modified Vlasov equation, in section \ref{section 4} we derived the related moments, i.e., continuity, momentum, and pressure/energy equations, which provide a fluid description of magnetized expanding plasmas. This description allowed us to characterize the physical meaning of the new effects introduced by the expansion. Namely, the EBM modifies the continuity equation written in the co-moving frame \eqref{eq: continuity eq EBM} by affecting the conservative form of the equation. 
The advantage is that within this framework there is no need to consider a "source" or a "sink" in the plasma description. 
Those effects appear naturally through the non-inertial forces in the modified Vlasov equation and represent the density decrease as the plasma moves away from its source (e.g., the Sun). Furthermore, the expanding momentum equation \eqref{eq: reduced momentum eq EBM} explicitly shows the non-inertial (fictitious) forces acting in the transverse directions of the plasma parcel as expected. Finally, in the EBM referential the conservative form of the pressure equation \eqref{eq: reduced pressure eq EBM} also changes. It should be noted that even if the heat flux is neglected [see Eq.~\eqref{e41.1}], the framework provides a way to describe the energetic losses due to the expansion (i.e., associated with the plasma cooling when it is expanding). This description allowed us to develop a general version of the pressure equation existing in the literature \cite{grappin.1993, Grappin.1996}, by explicitly developing it in its tensor form. This new result, which to our knowledge, has not been published before, enables to study more complex systems when an isotropic pressure is not sufficient to describe the plasma dynamics. The present results have also been tested in the limit case when there is no expansion (i.e., for a zero expanding velocity), and, as expected, we recovered the common equations given by Eqs. \eqref{e26} - \eqref{e28} not modified by the expanding effects. 

In Section~\ref{section 5},  we focused on developing an expanding (ideal) MHD description of the plasma, through the obtained moments and the usual approximations (e.g., neglecting electron's inertia, cold electron approximation, neglecting the Hall term, etc.). These results allow us to test the consistency and agreement between the modified expanding-Vlasov equation and the results in ideal-MHD, which in the literature were obtained only through a direct transformation of the inertial MHD equations to the co-moving frame \cite{grappin.1993, Grappin.1996}. It is not only these agreements that validate the developed framework, but also the physical meaning and correct interpretation of the non-inertial/fictitious term, which modifies the kinetic equation. This allows us to explicitly describe the expanding effects by modifying the conservative forms of the continuity, momentum, and energy equations. 
An accurate description and interpretation of the Vlasov equation in the co-moving frame is vital and substantial for a multi-scale agreement.

Our results should facilitate a series of applications and future investigations. Thus, the expanding moments \eqref{eq: continuity eq EBM}, \eqref{eq: reduced momentum eq EBM} and \eqref{eq: reduced pressure eq EBM} can be exploited, mainly through the pressure equation, to explore the modifications of the double adiabatic invariants given by the CGL equations ~\cite{Chew.1956}, which describes the evolution of parallel and perpendicular pressures when no heat flux is considered. Moreover, aiming to implement the kinetic effects in the MHD equations, as for a characterization of the Landau fluids \cite{Hunana_2018}, the expansion moments can also be used through, e.g., the heat flux term in equation \eqref{eq: reduced pressure eq EBM}. The advantage is that we can reproduce previous results in the EBM context, which enables to quantify the expansion effects by comparing them with the non-expanding results. Nevertheless, in the microscopic regime we can couple Vlasov equation with Maxwell equations to develop linear and quasi-linear theory for, e.g., spontaneous and induced emissions, and explore the effects of expansion at kinetic scales. Moreover, this work can be complemented if an accelerated expanding box is needed for the desired description \cite{Tenerani_2017}. 

The general theoretical formalism developed here can be applied to any astrophysical expanding plasma, from stellar to solar winds, or AGN jets to Coronal Mass Ejections (CMEs). As already motivated in the introduction, astrophysical plasmas are not static systems but are open and in continuous interaction with the environments. Therefore different approaches and considerations are needed in order to describe their dynamics. The underlying ideas of an EBM framework can provide essential support when a kinetic description of an expanding plasma is needed. In this paper, we aimed to introduce a stepping stone for future work. Even though here we have described non-relativistic plasma expansion at large (fluid) and small (kinetic) scales, a general relativity description can be employed through the coordinates transformations, allowing us to explore further applications of the developed framework. It thus stands out as one of the coordinate transformations by which methods of describing plasma physics in a co-moving framework can be adapted to any astrophysical scenario.
\vspace{6pt} 



\textbf{Authorcontributions: }{Conceptualization, S.E.-V., P.S.M, M.L.; methodology, S.E.-V., P.S.M, and M.L.; formal analysis, S.E.-V., P.S.M., M.L., and S.P.; investigation, S.E.-V., P.S.M., M.L., and S.P.; writing---original draft preparation, S.E.-V., P.S.M, M.L., and S.P.;  writing---review and editing, S.E.-V, P.S.M, M.L, and S.P. All authors contributed equally to this manuscript. All authors have read and agreed to the published version of the manuscript. 
}

\textbf{Funding: }{S. Echeverría-Veas is grateful to Agencia Nacional de Investigación y Desarrollo (ANID, Chile) for the National Doctoral Scholarship  N$^\circ$ 21211153. P. S. Moya thanks the support of FONDECyT grant No. 1191351 and the support of the Research Vice-rectory of the University of Chile (VID) through grant ENL08/23. M. Lazar and S. Poedts acknowledge support from the Ruhr-University Bochum and the Katholieke Universiteit Leuven. These results were also obtained in the framework of the projects G.0025.23N (FWO-Vlaanderen) and SIDC Data Exploitation (ESA Prodex-12).}

\textbf{Acknowledgments: }{We thank Dr. Felipe Asenjo for useful discussions in the modified Vlasov equation.}

\textbf{Conflicts of interest: }{The authors declare no conflict of interest.} 



\textbf{Abbreviations}{\\
The following abbreviations are used in this manuscript:\\

\noindent 
\begin{tabular}{@{}ll}
EBM & Expanding Box Model\\
MHD & Magnetohydrodynamic\\
VDF & Velocity Distribution Function\\
\end{tabular}
}

\appendix
\section*{Appendix A}
\label{section: appendix A}

For transforming all the physical quantities related to the EBM we align the $\hat{x}$ axis with the spherical radial coordinate. In this frame, the radial expanding velocity $\textbf{U}_0= U_0 \hat{r}$ read as
\begin{align*}
    \textbf{U}_0 = U_0 \hat{r} = U_0 \left( \frac{x'+R}{r}, \frac{a y'}{r }, \frac{a z'}{r}  \right)\,.
\end{align*}
The spherical radial coordinate $r$ can be written in terms of the new coordinates
\begin{align*}
    r &= \sqrt{x^2 + y^2 + z^2} = \sqrt{(x' + R)^2 + a^2 y'^2 + a^2 z'^2}\\
    &\approx R \left( 1 + \frac{x'}{R} + \frac{1}{2}\left( \frac{x'^2}{R^2} + \frac{y'^2}{R_0^2} + \frac{z'^2}{R_0^2} \right) \right)\,,
\end{align*}
for small parameters $x',y',z' \ll R$. Therefore, the expanding velocity in first order, read as
\begin{align*}
    \textbf{U}_0 = U_0 \left( \hat{x} + \frac{y'}{R_0} \hat{y} + \frac{z'}{R_0}\hat{z}\right)\,.
\end{align*}
Thus, temporal and spatial derivatives are
\begin{align*}
    \frac{\partial}{\partial t} &= \frac{\partial }{\partial t'} -  \textbf{U}_0 \cdot \nabla\\ & =  \frac{\partial }{\partial t'} - U_0 \left( \frac{\partial}{\partial x'}+  \frac{y'}{R}\frac{\partial}{\partial y'}+ \frac{z'}{R} \frac{\partial}{\partial z'} \right)\\
    &= \frac{\partial }{\partial t'} - \textbf{D}\cdot \nabla'\,,\\
    \nabla &=  \left( \frac{\partial}{\partial x'}, \frac{1}{a}\frac{\partial}{\partial y'} , \frac{1}{a}\frac{\partial}{\partial z'}\right)= \mathbb{A}^{-1} \cdot \nabla'\,,
\end{align*}
where
\begin{equation*}
        \textbf{D} = U_0 \left(1, \frac{y'}{R}, \frac{z'}{R} \right)\,,
      \quad   \mathbb{A} (t)= \begin{pmatrix}
    1 & 0 & 0 \\
    0 & {a} & 0 \\
    0 & 0 & {a}
    \end{pmatrix}\,.
\end{equation*}

On the other hand, the velocity in the co-moving frame in terms of the inertial variables
\begin{align*}
    v'_x &= \frac{d x'}{d t'} = \frac{d(x-R)}{d t} = v_x - U_0\,,\\
    v'_y &= \frac{d y'}{d t'} = \frac{d}{d t} \left(\frac{1}{a} y\right) = \frac{1}{a} v_y  - \frac{y'}{a} \frac{d a}{d t}  \,,\\ 
     v'_z &= \frac{d z'}{d t'} = \frac{d}{d t} \left(\frac{1}{a} z\right) = \frac{1}{a} v_z  - \frac{z'}{a} \frac{d a}{d t}  \,.\\ 
\end{align*}
Thus, velocity written in $S'$
\begin{align}
    \textbf{v} &= \left(v'_x + U_0, a v'_y + \frac{d a}{d t} y', a v'_z + \frac{d a}{d t} z'  \right) \nonumber\\
    & = \left( v'_x, a v'_y, a V'_z \right) + \frac{U_0}{R_0}\left( R_0,y' , z'\right)\nonumber\\
    &= \mathbb{A}\cdot \textbf{v}' + \textbf{U}_0 \,, \label{eq: Velocity in S'}
\end{align}
From that result we can also write the velocity gradient as
\begin{align*}
    \nabla_\textbf{v} &= \left(\frac{\partial}{\partial v_x} , \frac{\partial}{\partial v_y} , \frac{\partial}{\partial v_z}\right) = \left( \frac{\partial v'_x}{\partial v_x} \frac{\partial}{\partial v'_x}, \frac{\partial v'_y}{\partial v_y} \frac{\partial}{\partial v'_y}, \frac{\partial v'_z}{\partial v_z} \frac{\partial}{\partial v'_z}    \right) \\
    &=  \left(\frac{\partial}{\partial v'_x} , \frac{1}{a}\frac{\partial}{\partial v'_y} ,\frac{1}{a} \frac{\partial}{\partial v'_z}\right) = \mathbb{A}^{-1}\cdot \nabla_{\textbf{v}'}\,,
\end{align*}

Now we can write Vlasov's equation in the co-moving frame. On one side, the first term in Eq.~(\ref{eq: vlasov})
\begin{align}
    \frac{\partial f}{\partial t} = \frac{\partial f'}{\partial t'} -\left( \textbf{D} \cdot \nabla'\right) f' \,, \label{eq: 1st term vlasov}
\end{align}
where $f' = f(\textbf{r}', \textbf{v}', t')$. 
On the other hand, the second term of Eq.~(\ref{eq: vlasov})
\begin{align}
    \textbf{v} \cdot \frac{\partial f}{\partial \textbf{r}} &= \left[ \mathbb{A}\cdot \textbf{v}' + \textbf{U}_0\right]\cdot \left[\left(\mathbb{A}^{-1} \cdot \nabla'\right) f' \right] \nonumber\\ 
    &= \textbf{v}' \cdot \nabla' f' + \textbf{U}_0 \cdot \left[\left(\mathbb{A}^{-1} \cdot \nabla'\right) f' \right] \label{eq: 2nd term vlasov}
\end{align}
Adding the first and second term (Eqs. [\ref{eq: 1st term vlasov}] , [\ref{eq: 2nd term vlasov}]) from Vlasov's equation, the expressions remains the same but with the prime variables
\begin{align*}
    \frac{\partial f}{\partial t} + \textbf{v}\cdot \frac{\partial f}{\partial \textbf{r}} = \frac{\partial f'}{\partial t'} + \textbf{v}'\cdot \frac{\partial f'}{\partial \textbf{r}'}\,,
\end{align*}
due to
\begin{align*}
    \textbf{U}_0 \cdot \left[\left(\mathbb{A}^{-1} \cdot \nabla'\right) f' \right] - \left( \textbf{D} \cdot \nabla'\right) f' = 0\,.
\end{align*}
But it was also expected due to the first two terms in Vlasov equation are a total time derivative $\frac{d f}{d t}$. As we are working in a non-relativistic transformation $dt = dt'$, therefore $\frac{d f}{d t} = \frac{d f'}{d t'}$. Finally, we only need to write the acceleration $\textbf{a}$. In the inertial system this was equal to the Lorentz Force but as we are working in a non-inertial system this term will be different
\begin{align*}
    \textbf{a}= \frac{d \textbf{v}}{d t} = \frac{q}{m}\left( \textbf{E} + \frac{1}{c} \textbf{v}\times \textbf{B} \right)\,.
\end{align*}

With the results  from Eq.~(\ref{eq: Velocity in S'}) we can calculate the acceleration in $S'$
\begin{align}
    \frac{d \textbf{v}}{d t} &= \frac{d }{d t'}\left(\mathbb{A} \cdot \textbf{v}' + \textbf{U}_0 \right) \nonumber\\
    & = \mathbb{A} \cdot \frac{d \textbf{v}'}{d t'} + \frac{U_0}{R_0} \mathbb{T} \cdot \textbf{v}' \,, \label{eq: Acceleration in S'}
\end{align}
where $\mathbb{T}$ is a diagonal matrix that projects the vector in the perpendicular directions respect to the expansion
\begin{align*}
    \mathbb{T} = \begin{pmatrix}
    0 & 0 & 0 \\
    0 & 1 & 0 \\
    0 & 0 & 1
    \end{pmatrix}\,,
\end{align*}
and the first term of Eq.~\eqref{eq: Acceleration in S'} is the Lorentz force measured in the co-moving frame using field transformations \eqref{e2.14}
\begin{align}
    m \frac{d \textbf{v}'}{d t'} = q \textbf{E}' + \frac{q}{c} \left[\mathbb{A}\cdot \textbf{v}'\right]\times \textbf{B}' \,.
\end{align}
Therefore, the acceleration in the co-moving frame is given by
\begin{align}
    \textbf{a}' = \frac{q_\mu}{m_\mu} \mathbb{A}\cdot \left( \textbf{E}' + \frac{1}{c} \left[ 
 \mathbb{A}\cdot \textbf{v}'\right]\times \textbf{B}'\right) - \frac{U_0}{R_0} \mathbb{T}\cdot \textbf{v}'\,.
\end{align}

\section*{Appendix B}
\label{section: appendix B}

In this appendix we will develop in detail the momentum and pressure equation in the context of the EBM. For the first case we integrate Vlasov's equation \eqref{eq: Vlasov eq in S'} by $\int \textbf{v} d \textbf{v}$. The first term read as
\begin{align*}
    I_1 = \int \textbf{v} \frac{\partial f}{\partial t} d \textbf{v} = \frac{\partial }{\partial t} \int \textbf{v} f d \textbf{v} = \frac{\partial}{\partial t}\left( n \textbf{u}\right)\,.
\end{align*}

The second term
\begin{align}
    I_2 = \left[\int \textbf{v} \left( \textbf{v} \cdot \nabla\right) f d \textbf{v}\right]_i = \int v_i v_j \frac{\partial f}{\partial x_j} d \textbf{v} = \frac{\partial}{\partial x_j}\int v_i v_j f d \textbf{v}.
    \label{eq: I2_momentum}
\end{align}
For solving this integral we decompose the velocity between its mean $\textbf{u}$ and the fluctuations $\textbf{w}$
\begin{align*}
    \textbf{v} = \textbf{u} + \textbf{w}\,.
\end{align*}
Therefore
\begin{align*}
    v_i v_j = \left( u_i + w_i \right)\left( u_j + w_j\right) = u_i u_j + u_i w_j + w_i u_j + w_i w_j\,.
\end{align*}
Thus, in Eq.~\eqref{eq: I2_momentum}
\begin{align*}
    I_2 &= \frac{\partial}{\partial x_j} \int \left( u_i u_j + u_i w_j + w_i u_j + w_i w_j\right) f d \textbf{v}\\
    &= \frac{\partial}{\partial x_j} \int \left( u_i u_j +  w_i w_j \right) f d \textbf{v} = \frac{\partial}{\partial x_j} \left( u_i u_j n \right) + \frac{1}{m} \frac{\partial \mathbb{P}_{ij}}{\partial x_j}\,.
\end{align*}
Note that $\mathbb{P}$ is symmetric. Hence
\begin{align*}
    I_2 = \frac{\partial}{\partial x_j} \left( u_i u_j n \right) + \frac{1}{m} \frac{\partial \mathbb{P}_{ji}}{\partial x_j}\,.
\end{align*}
In vector notation
\begin{align*}
    I_2 = \nabla \cdot \left( \textbf{u} \textbf{u} n \right) + \frac{1}{m} \nabla \cdot \mathbb{P}\,,
\end{align*}
where the inner product between a vector $\textbf{V}$ and a tensor $\textbf{C}$ is defined as
\begin{align*}
    \left[\textbf{V} \cdot \mathbb{C} \right]_{i} = V_j C_{ji}\,,\\
    \left[ \nabla \cdot \mathbb{C} \right]_{i}= \frac{\partial C_{ji}}{\partial x_j}\,.
\end{align*}
The electric field integral
\begin{align*}
    I_3 &= \int E_i  \frac{\partial f}{\partial v_i} v_j \textbf{v} = E_i  \int \frac{\partial f}{\partial v_i} v_j d \textbf{v}\\
    &= - E_i  \int f \delta_{ij} \textbf{v} = - n  \left[\textbf{E}\right]_i\,.
\end{align*}

On the other hand, for the magnetic field
\begin{align*}
    I_4 &= \int {}\left\{\left[\mathbb{A}\cdot \textbf{v}\right]\times \textbf{B}\right\} \cdot \left[  \nabla_{\textbf{v}} f \right] \textbf{v} \textbf{v}\,,
\end{align*}
in index notation
\begin{align*}
    I_{4}&= \int \mathbb{A}_{ii} v_i B_j \epsilon_{ijk}  \frac{\partial f}{\partial v_k} v_l \textbf{v} \\
    &= - \mathbb{A}_{ii} B_j \epsilon_{ijk}  \int f \frac{\partial}{\partial v_k} \left( v_i v_l \right) \textbf{v}\\
    &= - n \mathbb{A}_{ii} B_j \epsilon_{ijk}  \left( \delta_{ik} u_l + u_i \delta_{lk} \right) \\
    &= - n \mathbb{A}_{ii} B_j \epsilon_{ijk}  u_i\,.
\end{align*}
In vector notation
\begin{align*}
    I_4 = - n  \left[\left( \mathbb{A} \cdot \textbf{u} \right) \times \textbf{B} \right]\,.
\end{align*}

Finally, the integral for this case is
\begin{align*}
    I_5 & = \frac{U_0}{R_0} \int \left( \mathbb{T} \cdot \textbf{v} \right) \cdot \left[\left(\mathbb{A}^{-1}\cdot \nabla_{\textbf{v}}\right) f \right] \textbf{v} d \textbf{v} 
     \\
     &=\frac{U_0}{a R_0}\int T_{ii} v_i  \frac{\partial f}{\partial v_i} v_j d \textbf{v}\\
     & = - \frac{U_0}{a R_0} \int \left( T_{ii} v_j + T_{ii} v_i \right) f d \textbf{v} \\
     & = - \frac{n U_0}{a R_0} \left[ \left( 2 \mathbb{I} + \mathbb{T} \right)\cdot \textbf{u}\right]_i \,,
\end{align*}
where $\mathbb{I}$ is identity. Using the continuity equation (\ref{eq: continuity eq EBM}) we obtained the expanding-momentum equation
\begin{align}
    \frac{\partial \textbf{u}}{\partial t}  + \left( \textbf{u} \cdot  \nabla\right) \textbf{u} = &\frac{q}{m}   \left[ \textbf{E} + \frac{1}{c} \left( \mathbb{A} \cdot \textbf{u} \right) \times \textbf{B} \right] \nonumber
    \\ &- \frac{1}{\rho} \nabla \cdot \mathbb{P}  - \frac{U_0}{a R_0} \mathbb{T}\cdot \textbf{u}\,.
\end{align}

For the pressure/energy equation we integrate Vlasov's equation by $\int \textbf{v} \textbf{v} d \textbf{v}$. Again, the first two terms of this equation are the same as in the non-expanding case. For a detailed derivation see \textcite{Hunana_2019} or \textcite{Webb_2022} (we will follow the same definitions). The first term from the Vlasov's equation
\begin{align}
    I_1 &= \int \textbf{v} \textbf{v} \frac{\partial f}{\partial t} d \textbf{v} 
     = \frac{\partial}{\partial t}\left(\textbf{u} \textbf{u} n + \frac{1}{m}\mathbb{P} \right)\label{eq: I_1 third moment}\,.
\end{align}

The second term
\begin{align*}
    I_2 = \int \textbf{v} \textbf{v} \left(\textbf{v}\cdot \nabla \right) f d \textbf{v}.
\end{align*}
Solving in index notation
\begin{align*}
    I_2 &= \int v_i v_j v_k \frac{\partial f}{\partial x_k} d \textbf{v}
    = \frac{\partial}{\partial x_k} \int v_i v_j v_k f d \textbf{v}\,.
\end{align*}
The non-zero terms from the integral
\begin{align*}
    I_2 = \frac{\partial}{\partial x_k} \int \left( u_i u_j u_k + u_i w_j w_k + w_i u_j w_k  + w_i w_j u_k + w_i w_j w_k \right) f d \textbf{v}
\end{align*}
\begin{align}
    = &\frac{\partial}{\partial x_k}\left[ n u_i u_j u_k + \frac{1}{m} u_i \mathbb{P}_{jk} + \frac{1}{m} u_j \mathbb{P}_{ik}  + \frac{1}{m} u_k \mathbb{P}_{ij}  + \frac{1}{m} \mathbb{Q}_{ijk} \right]\,. \label{eq: I_2 third moment all terms}
\end{align}
Note that the first and the last two terms from \eqref{eq: I_2 third moment all terms} can be identified as
\begin{align}
    \frac{\partial}{\partial x_k} \left[ n u_i u_j u_k + \frac{1}{m} u_k \mathbb{P}_{ij}  + \frac{1}{m} \mathbb{Q}_{ijk} \right]
    = \nabla \cdot \left[ n \textbf{u} \textbf{u} \textbf{u} + \frac{1}{m} \textbf{u} \mathbb{P} + \frac{1}{m} \mathbb{Q} \right]_{ij}\,. \label{eq: I_2 third moment}
\end{align}
The remaining terms will be treated later. On the other hand, the electric field integral
\begin{align*}
    I_3 = \int \textbf{v} \textbf{v} \left[\textbf{E} \cdot  \nabla_{\textbf{v}} f \right] d \textbf{v} \,.
\end{align*}
In index notation
\begin{align*}
    I_3 &= \int v_i v_j E_k  \frac{\partial f}{\partial v_k} d \textbf{v} = - E_k  \int \frac{d }{d v_k} \left(v_i v_j \right) f d \textbf{v}\\
    &= - n E_{ii}  u_j - n E_{jj}  u_i\,.
\end{align*}
In tensor notation
\begin{align}
     I_3= - n \left[\textbf{u} \textbf{E}   \right]^s\,, \label{eq: I_3 third moment}
\end{align}
where the supra-index $s$ represents a symmetric operator that acts on a matrix $\mathbb{C}$ as $\mathbb{C}^s = \mathbb{C} + \mathbb{C}^T$ or $C_{ij}^s = C_{ij} + C_{ji}$. For the magnetic field integral
\begin{align*}
    I_4 = \int \textbf{v} \textbf{v} \left[ \left(\mathbb{A} \cdot \textbf{v} \right) \times \textbf{B}\right] \cdot  \nabla_{\textbf{v}} f d \textbf{v}\,.
\end{align*}
In index notation
\begin{align*}
    I_4 &= \int v_i v_j \mathbb{A}_{kk} v_k B_l \epsilon_{klm} \frac{\partial f}{\partial v_m} d \textbf{v}\\
     &= - \mathbb{A}_{kk}  B_l \epsilon_{klm} \int \frac{\partial}{\partial v_m} \left(v_i v_j v_k \right) f d \textbf{v} \\
     &= - \mathbb{A}_{kk}  B_l \epsilon_{kli}\left(n u_j u_k + \frac{1}{m} \mathbb{P}_{jk} \right) - \mathbb{A}_{kk}  B_l \epsilon_{klj}\left(n u_i u_k + \frac{1}{m} \mathbb{P}_{ik} \right)\,.
\end{align*}
In tensor notation
\begin{align}
    I_4 = &- n \left[  \left( \left(\mathbb{A} \cdot \textbf{u}\right)\times \textbf{B} \right) \textbf{u}\right]^s 
    +\frac{1}{m} \left[  \textbf{B}\times \left(\mathbb{A}\cdot \mathbb{P} \right) \right]^s\,, \label{eq: I_4 third moment}
\end{align}
where the cross product between a vector and a tensor is defined as
\begin{align*}
    \left[ \textbf{B} \times \mathbb{P} \right]_{ik} = \epsilon_{ijl}B_j \mathbb{P}_{lk}\,.
\end{align*}
Finally, the expanding integral in this case read as
\begin{align*}
    I_5 & = \frac{U_0}{R_0} \int \left( \mathbb{T} \cdot \textbf{v}' \right) \cdot \left[\left(\mathbb{A}^{-1}\cdot \nabla_{\textbf{v}'}\right) f \right] \textbf{v} \textbf{v} d \textbf{v} \\
   & = \frac{U_0}{a R_0} \int T_{ii} v_i  \frac{\partial f}{\partial v_i} v_j v_k d \textbf{v}\\
    & = - \frac{U_0}{a R_0} \left( 2 \int v_j v_k f d \textbf{v} +  T_{jj} \int v_j v_k f d \textbf{v} 
     +  T_{kk} \int v_k v_j f d \textbf{v}\right)\,.
\end{align*}
For solving those integrals, we only need to decompose the velocity in terms of the mean velocity and its fluctuations $\textbf{v} = \textbf{u} + \textbf{w}$, obtaining 
\begin{align*}
    I_5 = - \frac{ 2 U_0}{a R_0} \left(\textbf{u} \textbf{u} n + \frac{1}{m} \mathbb{P} \right)- \frac{U_0}{a R_0} \left[  \mathbb{T}\cdot \left( \textbf{u}\textbf{u} n + \frac{1}{m}\mathbb{P} \right)  \right]^s\,.
\end{align*}

We can reduce the equations if we use the continuity and momentum equation. Note that from the first terms of Eqs. (\ref{eq: I_1 third moment}), (\ref{eq: I_2 third moment})
\begin{align*}
    \frac{\partial}{\partial t}\left[\textbf{u} \textbf{u}n \right]_{ij} + \nabla\cdot\left[ \textbf{u} \textbf{u} \textbf{u} n \right]_{ij} 
     &= \left[ \left( \frac{\partial u_i}{\partial t} + u_k \frac{\partial u_i}{\partial x_k} \right) u_j n\right]^s \\
     &+ u_i u_j \left[ \frac{\partial n}{\partial t} + \frac{\partial u_k}{\partial x_k} n + u_k \frac{\partial n}{\partial x_k} \right]\,.
\end{align*}
The first parenthesis can be written in terms of the momentum equation, and the second in terms of the continuity equation
\begin{align}
    &\frac{\partial}{\partial t}\left[\textbf{u} \textbf{u}n \right] + \nabla\cdot\left[ \textbf{u} \textbf{u} \textbf{u} n \right] \nonumber \\
    &= \frac{q n}{m} \left[  \left\{ \textbf{E} + \frac{1}{c} \left( \mathbb{A} \cdot \textbf{u} \right) \times \textbf{B} \right\} \textbf{u} \right]^s - \frac{1}{m} \left[ \textbf{u} \nabla \cdot \mathbb{P}\right]^s \nonumber \\
    &\;\;\;\; - \frac{U_0}{a R_0} \left[\textbf{u} \mathbb{T} \cdot \textbf{u} \right]^s - \frac{2 n U_0}{a R_0} \textbf{u} \textbf{u}\,. \label{eq: reducing pressure}
\end{align}
Note that the first term from this equation is the same as in Eq.~(\ref{eq: I_3 third moment}) and the first one from Eq.~(\ref{eq: I_4 third moment}), but with opposite sign. Finally, recall that from Eq.~(\ref{eq: I_2 third moment all terms}) we still need to work with second and third terms. From Eq.~(\ref{eq: reducing pressure})
\begin{align*}
    \frac{\partial}{\partial x_k} \left( u_i \mathbb{P}_{jk} + u_j \mathbb{P}_{ik} \right) - \left[ \textbf{u} \nabla \cdot \mathbb{P}\right]_{ij}^s 
     & = \mathbb{P}_{ik} \frac{\partial u_j}{\partial x_k} + \mathbb{P}_{jk} \frac{\partial u_i}{\partial x_k} \\
    & = \left[ \mathbb{P} \cdot \nabla \textbf{u} \right]_{ij}^s\,.
\end{align*}

Finally, the expanding pressure equation reads as
\begin{align}
    &\frac{\partial \mathbb{P}}{\partial t} + \nabla \cdot \left( \textbf{u} \mathbb{P} + \mathbb{Q} \right) \nonumber + \left[ \mathbb{P} \cdot \nabla \textbf{u} \right]^s + \frac{q}{m c} \left[  \textbf{B} \times 
 \left( \mathbb{A} \cdot \mathbb{P} \right) \right]^s \nonumber \\
    &= - \frac{2 U_0}{a R_0} \mathbb{P} - \frac{U_0}{a R_0} \left[ \mathbb{T} \cdot \mathbb{P}\right]^s\,.
\end{align}

\printbibliography

\end{document}